# Discrete optics in optomechanical waveguide arrays

Xinbiao Xu,[†,§,⊥] Linhao Ren,[†,⊥] Lei Shi,[†,*] and Xinliang Zhang[†]

[†]Wuhan National Laboratory for Optoelectronics, Huazhong University of Science and Technology, Wuhan 430074, China

[§]CAS Key Laboratory of Quantum Information, University of Science and Technology of China, Hefei 230026, China

[*]Corresponding Author: lshi@hust.edu.cn

**ABSTRACT:**

Propagation properties of light in optomechanical waveguides arrays (OMWAs) are studied for the first time, to the best of our knowledge. Due to the strong mechanical Kerr effect, the optical self-focusing and self-defocusing phenomena can be realized in the arrays of subwavelength dielectric optomechanical waveguides with the milliwatt-level incident powers and micrometer-level lengths. Compared with the conventional nonlinear waveguide arrays, the required incident powers and lengths of the waveguides are decreased by five orders of magnitude and one order of magnitude, respectively. Furthermore, by adjusting the deformation of the nanowaveguides through a control light, the propagation path of the signal light in the OMWA can be engineered, which could be used as a splitting-ratio-tunable beam splitter. This work provides a new platform for discrete optics and broadens the application of integrated optomechanics.

Over the past few decades, photonic integrated circuits have been extensively studied. With the extraordinary progress of micro/nano-processing and material growth technologies, low-loss waveguides and high-quality-factor microcavities made by different dielectric materials, such as silicon nitride, aluminum nitride and lithium niobite, have been realized in photonic integrated circuits.[1-5] Benefiting from the progress, high-performance soliton frequency microcombs have been realized by controlling the balance of dispersion and Kerr nonlinearity in the microcavities.[6-9] Similar to the temporal solitons generated in the microcavities, the arrays or lattices of evanescently coupled nonlinear waveguides are excellent platforms for discrete optical dynamics research.[10] For example, the spatial optical solitons, the diffraction management, and the Bloch oscillation of light have been observed in the coupled waveguide arrays.[11-17] The spatial soliton is a self-trapped wavepacket with an unchanged profile attributed to the balance between diffraction and Kerr nonlinearity in the coupled waveguide array. However, limited by the relatively weak intrinsic Kerr coefficients of conventional dielectric materials, the discrete spatial solitons are usually observed in the weakly coupled waveguide arrays. Consequently, the sizes of the waveguides are always large (i.e., micrometer-level width and height, and millimeter-level length) and hundred-watt-level optical powers are needed.[10, 11, 16, 18-20] High optical powers will cause some undesired results, such as the additional losses due to the nonlinear absorption.[21] Besides, the tight confined surface plasmon polariton modes in metal-dielectric composites can greatly enhance the effective nonlinear coefficient.[22] Therefore, the sizes of the waveguides required for the spatial soliton generation can be significantly decreased in the metal-dielectric waveguide or graphene sheet arrays.[23-25] However, the metal-dielectric waveguide or graphene sheet arrays are difficult to be fabricated and the propagation losses of the plasmonic modes are very large.[26, 27]

Recently, an extremely strong mechanical Kerr effect induced by the optical gradient force

(OGF) gets increasing attention.[28-37] The OGF between the suspended waveguide and the substrate can lead to the mechanical deformation of the waveguide. Owing to the deformation, the effective refractive index ($n_{eff}$) of the guided mode changes, which is called mechanical Kerr effect.[38] The mechanical Kerr coefficient can be several orders of magnitude larger than the conventional optical Kerr coefficient,[39] which means that light field manipulation can be greatly enhanced. As the stronger OGF leads to the larger waveguide deformation, several optomechanical structures have been proposed to enhance the OGF, which are based on metamaterials,[40] metal-dielectric waveguides[41] and graphene.[42,43] Due to the strong mechanical Kerr effect, OGFs have been used for tunable directional couplers,[44] tunable microcavities,[45] dispersion engineering,[46] synchronization of nanomechanical oscillators and optical nonreciprocal transmission.[47,48]

In this work, the propagation properties of light in the subwavelength dielectric waveguide arrays with the mechanical Kerr effect are studied. Our work shows that without the plasmonic modes to enhance the nonlinearity, typical discrete optical phenomena such as the self-focusing and self-defocusing can be realized in the strongly coupled optomechanical waveguide arrays (OMWAs) with the microwatt-level incident powers and micrometer-level lengths.

**Discretized light behavior in OMWAs**

The photonic system we proposed and studied is shown in Figure 1, which consists of 17 suspended and double-clamped nanobeams (DCB). All the DCBs are identical and equidistant single-mode rectangular nanowaveguides. The material of the DCB is silicon nitride ($Si_3N_4$) with the refractive index $n = 2$ at the wavelength of 1.55 μm and the substrate is made of silica ($SiO_2$). As shown in Figure 2a, the waveguide cross section is 800 nm × 400 nm and the length of the suspended waveguide is $L = 600$ μm. The horizontal separation between the adjacent waveguides is $S=1$ μm. The initial gap between the suspended waveguide and the substrate is $g = 300$ nm.

When light propagates through the DCB, the OGF density (N/m/W) between the waveguide and the substrate can be calculated by $f_n(g) = \frac{1}{c}\frac{\partial n_{eff}}{\partial g}$,[40] where $c$ is the light speed in vacuum. As shown in Figure 2b, $n_{eff}$ is inversely proportional to the gap $g$. Therefore, the OGF is attractive ($f_n(g) < 0$) and it is inversely proportional to the gap $g$ as well. The attractive OGF bends the waveguide towards the substrate. The deflection of the waveguide is dependent on the distribution of the optical power along the waveguide, which will change the mode field distribution and $n_{eff}$. Hence the coupling coefficient between the adjacent waveguides and the phase of light (i.e. $\varphi = \varphi_0 + k_0 \cdot n_{eff} \cdot L$) are related to the optical power in the waveguide, which is similar to the conventional optical Kerr effect. Figure 2c describes the electric field distributions of the eigenmodes in the two coupled waveguides under different deflection conditions. In the top two pictures of Figure 2c, $g_1 = g_2$, this means that it is a symmetric directional coupler. However, as shown in the bottom two pictures of Figure 2c, there is the deformation-induced mode mismatching. The coupler becomes asymmetric and the coupling between the two waveguides is weaker. In order to analyze the light propagation in the OMWA, we numerically calculate the coupling coefficient between the two waveguides versus $g_1$ and $g_2$ in Figure 2d. It can be found that there is a maximum coupling coefficient when $g_1 = g_2 = 125$ nm.

Although the light coupling between the adjacent waveguides means that there is the horizontal optical force between the adjacent waveguides, the horizontal separation $S$ is relatively large so that the horizontal optical force between the adjacent waveguides is ~$10^{-6}$ nN/μm/mW. It is at least two orders of magnitude smaller than the vertical optical force between the waveguide and the substrate. Hence in the following simulations, the horizontal optical force is neglected and only the power coupling between the adjacent waveguides is considered. We consider a Gaussian beam as the incident beam. The light propagation in the OMWA is described by the coupled mode

equations:

$$\frac{dA_n(z)}{dz} = i\kappa_{n,n+1}(z)A_{n+1}(z)e^{-i2\delta_{n,n+1}(z)} + i\kappa_{n,n-1}(z)A_{n-1}(z)e^{-i2\delta_{n,n-1}(z)} \quad (1)$$

where $A_n(z)$ is the slowly-varying complex amplitude of the mode field in the *n-th* waveguide, $\kappa_{i,j}(z)$ is the coupling coefficient between the *i-th* and *j-th* waveguides, and $\delta_{i,j}(z) = (\beta_i(z) - \beta_j(z))/2$ represents the detuning of the propagation constants between the *i-th* and *j-th* waveguides. Besides, the waveguide deflection $u(z)$ is determined by Euler Bernoulli beam theory with the boundary conditions of $u(0) = u'(0) = u(L) = u'(L) = 0$,[28]

$$EI\frac{d^4u(z)}{dz^4} = f_n(z)P(z) \quad (2)$$

where $E = 300$ GPa is the Young's modulus of $Si_3N_4$,[49] $I = w^3h/12$ is the area moment of inertia of the DCB, $P(z)$ is the optical power distribution along the DCB, and $f_n$ is a function of $z$ as well. Once the initial optical field distribution in the OMWA and the incident power are given, the optical force distribution and the induced waveguide deflection can be calculated. The deflection will reversely influence the optical field distribution, leading to a new optical force distribution. Therefore, after certain iterative loops, a stable optical field distribution and waveguide deflection will be obtained.

The full width at half maximum of the incident Gaussian beam is set to be 3 μm. When the Gaussian beam is incident normally to the waveguides, the optical power distributions in the OMWA under the signal powers of 0.1 mW, 0.8 mW, and 1.293 mW are shown in Figure 3a-c, respectively. Figure 3d-f show the corresponding distributions of variation of the effective refractive index ($\Delta n_{eff}$) in the OMWA. When the signal power is 0.1 mW, the light beam spreads over more and more waveguides as it propagates, due to the coupling between the adjacent waveguides. In this case, the OGF is so small that the deflection of the waveguide can be neglected, and thus $\Delta n_{eff}$ of the waveguides are almost the same, as shown in Figure 3d. In other words, the mechanical Kerr effect is weak. When the signal power continues to increase, the deflection of the waveguide will be larger, but it is different for each waveguide because of the different optical power distributions along the OMWA. When the signal power is 0.8 mW, according to the optomechanical coupling, the $n_{eff}$ distribution of the OMWA will look like a "convex lens" at the final steady state. As a result, the light beam propagating through the OMWA exhibits the self-focusing effect as depicted in Figure 3b, e. When the signal power increases to 1.293 mW, the confinement of the light beam is stronger, and most of the optical field is confined in a few waveguides and the deflection of the central waveguide is much larger than those of the other waveguides. The maximum $\Delta n_{eff}$ can reach 0.002 as illustrated in Figure 3c, f. For the conventional optical Kerr effect, a power of 6.83 kW is needed to get the same $\Delta n_{eff}$. From this point of view, the mechanical Kerr coefficient of the OMWA is six orders of magnitude larger than the conventional optical Kerr coefficient. Besides, as for the optical Kerr effect, the optical field only changes the refractive index where the optical field is located. However, the mechanical Kerr effect is nonlocal so that the OGF at a certain position can lead to the deformation of the whole waveguide, and then $n_{eff}$ of the whole waveguide changes. Therefore, limited by the deformation manner of the waveguide, the spatial soliton is difficult to be obtained in the OMWA, but the strong self-focusing effect with a milliwatt-level incident power and micrometer-level waveguide length is realized, which is extremely challenging in the conventional nonlinear waveguide systems. It is important to note that, because of the complex spatial distribution of the mode field of the freestanding waveguide, the coupling efficiency does not change monotonously with the increasing of $g_2$ ($g_1$) for fixed $g_1$ ($g_2$), as shown in Figure 2d. There is a threshold for the signal power, where the coupling strengths at the position around $z = 300$ μm between the central waveguide and its adjacent waveguides get the maximum values. Once the signal power is larger than the threshold (about 1.795 mW), the deflection of the central waveguide will be larger, but the coupling coefficients decrease. Therefore, the optical field initially distributed in the other waveguides will concentrate to the central waveguide, and the deflected central waveguide will touch the substrate suddenly.

In this system, the coupling between the adjacent waveguides is a form of discrete diffraction, which is described by the diffraction coefficient $D_{n,n+1} = -2\kappa_{n,n+1}(z) S^2 \cos(k_x S)$, where $k_x$ is the x component of the wave vector. $k_x S$ represents the phase difference between the adjacent waveguides induced by the incident angle of the light beam. Accordingly, the magnitude and sign of the diffraction coefficient can be controlled, which cannot be achieved in a homogenous medium. The diffraction is normal in the range of $|k_x S| < \pi/2$. In the positive Kerr nonlinearity system, when the Kerr effect compensates the normal diffraction, the spatial soliton appears. With the continuous enhancement of the Kerr effect, there is the self-focusing of the light beam, as shown in Figure 3. The diffraction becomes anomalous in the range of $\pi/2 < |k_x S| \leq \pi$. Therefore, by introducing a relative phase shift between the adjacent waveguides at the input port, the light propagation properties will be different. Here, by adjusting the incident angle to about 22.8 degrees, we introduce a $\pi$ phase shift between the adjacent waveguides ($k_x S = \pi$), and then the diffraction becomes anomalous. Under the lower incident power, the light beam in the OMWA broadens as well, as shown in Figure 4a. However, when the incident power increases, the output light beam does not focus as it is normally incident but rather spreads and becomes significantly wider, as shown in Figure 4b, c. This is exactly opposite to the self-focusing effect, and called self-defocusing.

As shown above, the spatial distribution of $n_{eff}$ can be controlled by changing the optical power and the incident angle. Accordingly, the propagation path of the signal light can be controlled by a control light with a milliwatt-level power, which means that there is strong nonlinear interaction between the signal light and the control light. In order to describe the phenomenon, the signal light with a power of 1 μW is incident under the condition of $|k_xS| = \pi/2$ by adjusting the incident angle to about 11.7 degrees. As depicted in Figure 5a, the signal light can cross the OMWA without diffraction ($D_{n, n+1} = 0$), which means that the width of the light beam remains unchanged. In this case, the system works in the linear regime. When the Gaussian control light is incident normally to the central waveguide of the array, the strong OGF causes the waveguides to deform and a new $n_{eff}$ distribution forms in the OMWA. When the control light is 1.31 mW, $n_{eff}$ of the central waveguide is much larger than those of the other waveguides, which causes a portion of the signal light to be reflected at the central waveguide as shown in Figure 5b, e. When the control power continues to increase, a larger portion of the signal light will be reflected, which is shown in Figure 5c, f. Based on this mechanism, theoretically, we can obtain a beam splitter which possesses an arbitrary power splitting ratio by adjusting the control power. Figure 6 depicts the output power distributions of the signal light under different control powers. The waveguides in the yellow region are set as one port of a beam splitter, and the other waveguides are set as the other port. The result indicates that this structure can be used as an all-optical tunable beam splitter. Due to the large mechanical Kerr coefficient, a low control power of about 1.795 mW is needed to reflect 90% of the signal light.

**CONCLUSION**

We proposed an optomechanical discrete system and investigated the light propagation properties in the system. Because of the strong mechanical Kerr effect, when the light beam was incident normally to the waveguides, the self-focusing effect of the light beam is realized. However, by slightly changing the incident angle, the condition of anomalous diffraction is met and the self-defocusing is also realized. It is worth noting that these phenomena are achieved under the milliwatt-level incident powers and micrometer-level waveguide lengths, which are five orders of magnitude lower and one order of magnitude smaller than those in the conventional nonlinear waveguide arrays, respectively. In addition, we also proposed the application of a tunable beam splitter based on this system. Our work exhibits the potential of integrated optomechanics in strong nonlinear optical interaction and gives us a new platform to study discrete optics. Besides, because of the nonlocal property of the mechanical Kerr effect, the propagation properties of light are also different from those in the conventional nonlinear waveguide arrays.

## AUTHOR INFORMATION

Corresponding Author

*Email: lshi@hust.edu.cn

## Author Contributions

[⊥]X.B.X. and L.H.R. contributed equally to this work. L.S., X.B.X. and L.H.R. conceived the idea. X.B.X. and L.H.R. performed the simulations. L.S. supervised the project. All authors discussed the results and reviewed the manuscript.

## Notes

The authors declare no competing financial interest.

## ACKNOWLEDGMENTS

This work was supported by the National Natural Science Foundation of China (91850115, 11774110, 11947234), the Fundamental Research Funds for the Central Universities (HUST: 2019kfyXKJC036, 2019kfyRCPY092), and the Open Fund of IPOC (BUPT) (IPOC2019A012).

**Figure 1.** Schematic of the array of subwavelength dielectric optomechanical waveguides.

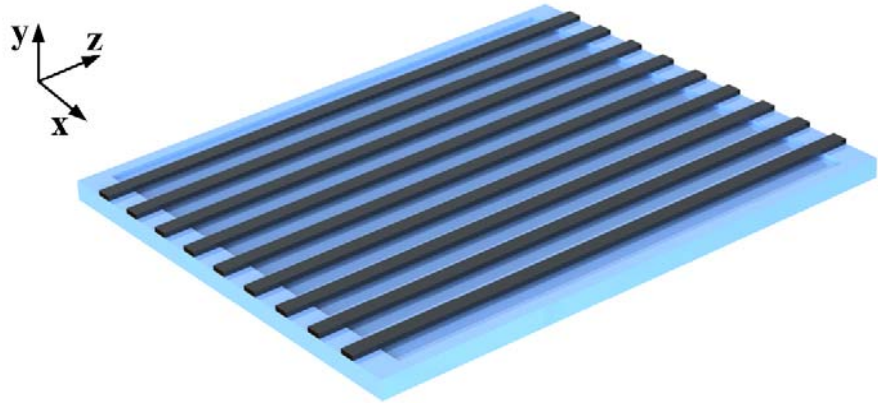

**Figure 2.** (a) Cross sections of two coupled optomechanical waveguides. (b) Effective refractive index and OGF density of the suspended waveguide versus the gap. (c) Electric field distributions of the eigenmodes of the two coupled waveguides under different deflection conditions. (d) Coupling coefficient between the two waveguides versus the gaps.

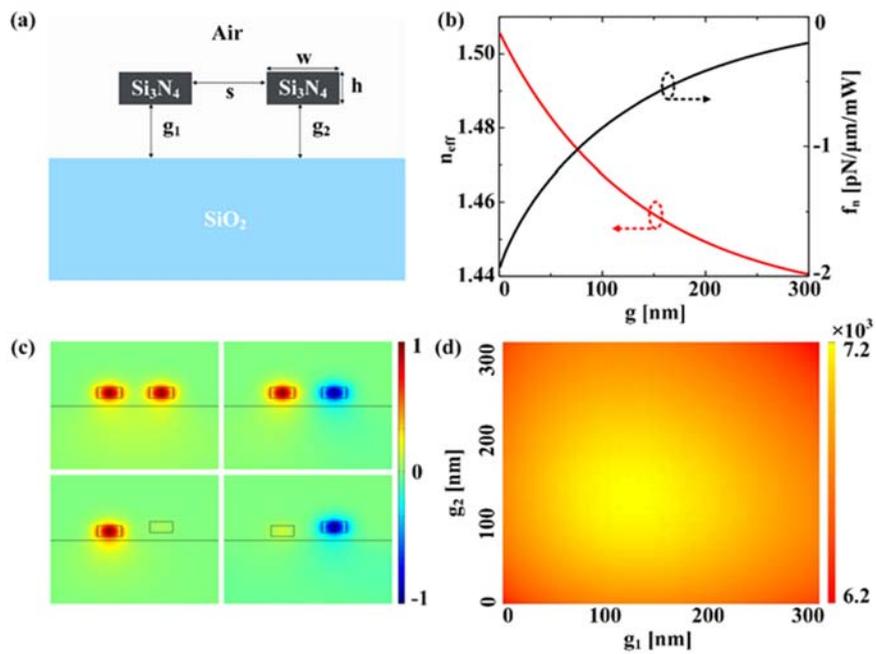

**Figure 3.** Self-focusing of light in the OMWA. (a-c) Optical power distributions in the OMWA under the signal powers of 0.1 mW, 0.8 mW, 1.293 mW, respectively. (d-f) Corresponding $\Delta n_{eff}$ distributions in the OMWA.

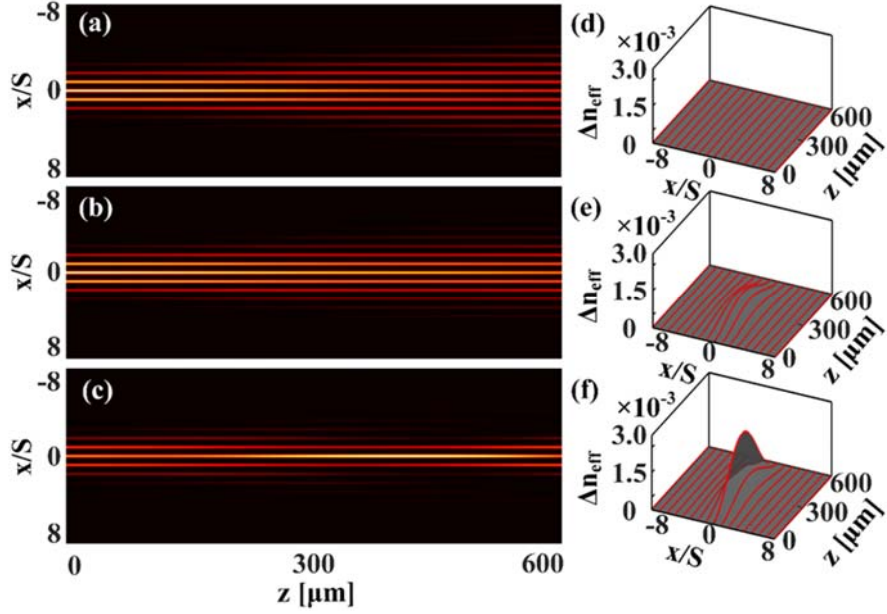

**Figure 4.** Self-defocusing of light in the OMWA. (a-c) Optical power distributions in the OMWA with an incident angle of 22.8 degrees under the signal powers of 0.1 mW, 1.293 mW, 2.8 mW, respectively. (d-f) Corresponding $\Delta n_{eff}$ distributions in the OMWA.

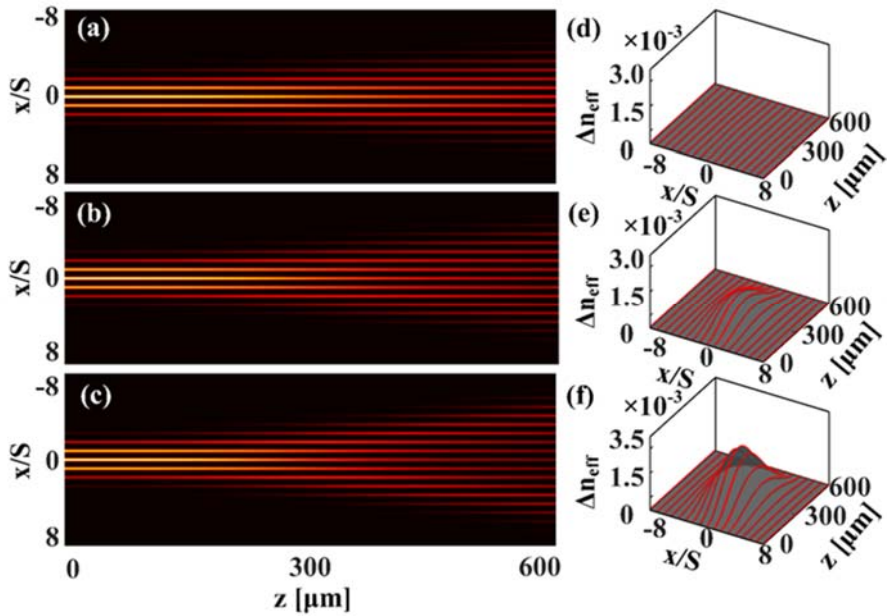

**Figure 5**. Reflection of light in the OMWA. Power distributions of the signal light in the OMWA under an incident angle of 11.7 degrees (a) without the control light, and (b,c) with the control light of 1.31 mW and 1.795 mW, respectively. (d-f) Corresponding $\Delta n_{eff}$ distributions in the OMWA.

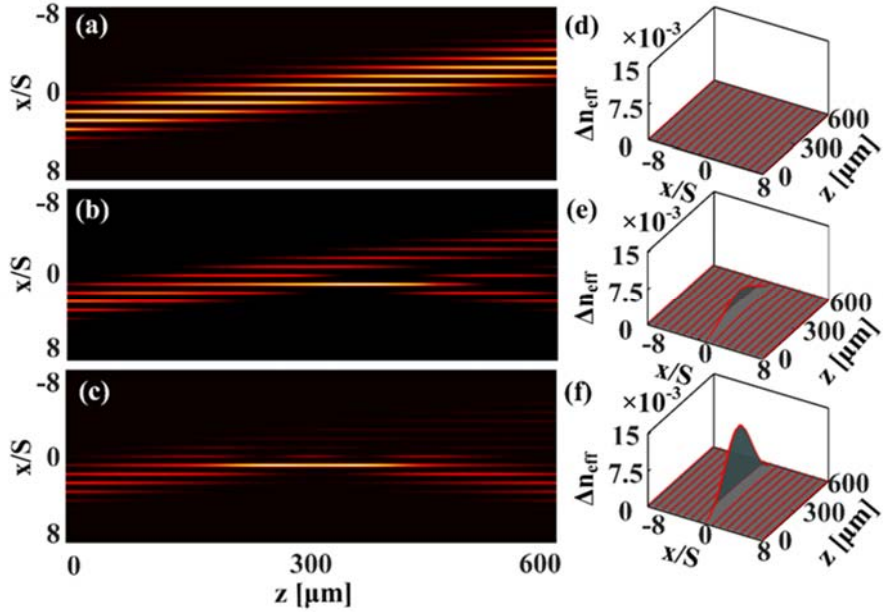

**Figure 6**. Output power distributions of the signal light under different control powers and its application for tunable beam splitting.

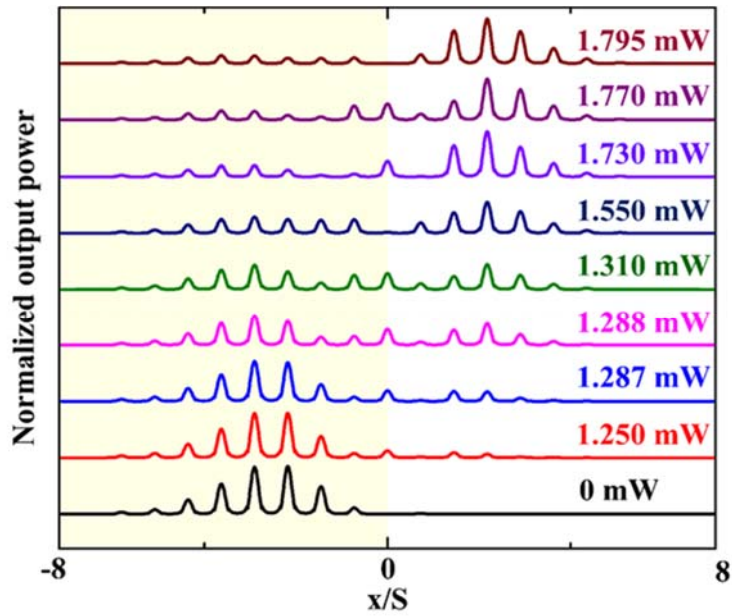